\documentclass{article}
\usepackage{amssymb}  \usepackage{amsmath} \usepackage{graphicx} 
\usepackage{subfigure}

\usepackage{fancyhdr}

\usepackage{calc} \usepackage{color} 
\newlength{\defaultparindent} 
\newlength{\anchocelda} 
 \newlength{\otroancho}
\newcommand{\n}{\rule{\anchocelda}{\anchocelda}}
\newcommand{\x}{\makebox[\anchocelda]{\rule{0mm}   {\anchocelda}}}
\newcommand{\y}{{\color{red}\n}} \newcommand{\w}{{\color{blue}\n}}

\begin{document}
\title{On Memory and Structural Dynamism in\\ Excitable Cellular Automata with Defensive Inhibition}
\date{}
\author{Ram\'on Alonso-Sanz$^{1}$ \and Andrew Adamatzky$^{2}$}
\maketitle

\begin{centering}
$^{1}$ ETSI Agr\'{o}nomos (Estad\'\i{}stica), C.Universitaria. 28040, Madrid\\
{\tt ramon.alonso@upm.es}\\
$^{2}$ Faculty of Computing, Engineering and Mathematical Sciences, \\ University of the West of England, Bristol, UK\\
 {\tt andrew.adamatzky@uwe.ac.uk}\\
\end{centering}


\abstract{Commonly studied cellular automata are memoryless and have fixed topology of connections 
between cells. However by allowing updates of links and short-term memory in cells we may 
potentially discover novel complex regimes of spatio-temporal dynamics. Moreover by adding memory and 
dynamical topology to state update rules we somehow forge elementary but non-traditional models of 
neurons networks (aka neuron layers in frontal parts). In present paper we demonstrate how this can be done
on a self-inhibitory excitable cellular automata. These automata imitate a phenomenon of 
inhibition caused by hight-strength stimulus: a resting cell excites if there are one or two excited 
neighbors, the cell remains resting otherwise.  We modify the automaton  by allowing cells to have 
few-steps memories, and make links between neighboring cells removed or generated depending on states of the
cells.


\pagestyle{fancy}
 
\lhead{\footnotesize Alonso-Sanz~R. and Adamatzky A. On Memory and Structural Dynamism in Excitable Cellular Automata with Defensive Inhibition. I. J. Bifurcation and Chaos  18 (2008) 527--539.}
\chead{}
\rhead{}

\section{Cellular Automata and their excitable species}

Cellular Automata (CA) are discrete, spatially explicit extended dynamic systems. A CA system is composed of adjacent cells or sites
arranged as a regular lattice, which evolves in discrete time steps. Each cell is characterized by an internal state whose value
belongs to a finite set. The updating of these states is made simultaneously according to a common local transition rule involving only a neighborhood of each cell \cite{ILA1},\cite{WOL1}. Thus, if $\sigma ^{(T)}_{i}$ is taken to denote the value of
cell \textit{i} at time step $T$, the site values evolve by iteration of the mapping: $\sigma ^{(T+1)}_{i} = \phi\Big( \sigma^{(T)}_{j}\in \mathcal{
N} _{i} \Big)$ , where $\phi$ is an arbitrary function which specifies
the cellular automaton \textit{rule} operating on the neighborhood $\mathcal{N}$ of the cell \textit{i}.

In {\it totalistic} rules the value of a site depends only on the sum of the values of its  neighbors, and not on their individual values: $\sigma^{(T+1)}_{i} =\phi \Big( \displaystyle\sum_{j\in\mathcal{N}_{i}}\sigma^{(T)}_{j}\Big)$ .

This paper deals with two-dimensional semi-totalistic excitable CA, where every cells take three states:
resting $0$, excited $1$ and refractory $2$. State transitions from excited to refractory and from refractory to 
resting are unconditional, they take place independently on cell's neighborhood state:
$\sigma^{(T)}_{i}=1 \rightarrow \sigma^{(T+1)}_{i}=2$, $\sigma^{(T)}_{i}=2 \rightarrow \sigma^{(T+1)}_{i}=0$.

In representing excitation rule we adopted Pavlovian phenomenon of defensive inhibition: when strength 
of stimulus applied to some parts of nervous system exceeds certain limit the system `shuts down', this can be
naively interpreted as an inbuilt protection of energy loss and exhaustion. To simulate the phenomenon of 
defensive inhibition we adopt interval excitation rules, developed in \cite{adamatzky_book_2001},
and put as a resting cell becomes excited only if one or two of its neighbors are excited.
If more then two neighbors excited the defensive inhibition comes into action and prevents the cell from 
excitation: $\sigma^{(T)}_{i}=0 \rightarrow \sigma^{(T)}_{i}=1$ if   $\displaystyle\sum_{j\in\mathcal{N}_{i}}\big(\sigma^{(T)}_{j}=1\big) \in \{1,2\}$. 

An example of simple development is shown in Fig.~{\ref{fig:sequence} where configurations of defensive-inhibition CA starting from an excited singleton, and evolving up to $T=15$; there, as well as Sect.~2 and initially in the remaining, the Moore neighborhood (eight neighbors 
plus central cell) is adopted.

The defensive inhibition rule was studied in \cite{adamatzky_holland} amongst other interval excitation rules in a context of 
morphological and dynamical complexities. We demonstrate there that the rule exhibits quite a low morphological complexity --- it 
is amongst rules with lowest number of different cluster (i.e. clusters of excited and refractory states) sizes. However CA governed 
by the rule exhibits longest transient periods. Calculating Langton's $\lambda$ parameter we also shown that the rule stays toward
the middle of the rule phases space, and occupies general position of rules with complex behavior. A contradiction between morphological, 
i.e. {\emph a posteriori}, and function-based $\lambda$, i.e. \emph{a priori}, measurements of complexity sounds alarming. Moreover, 
the closest to the defensive  inhibition rule -- in the interval excitation universe -- is the interval $[2,2]$ rule is a `kingdom of complexity 
and universality', as well demonstrated in \cite{adamatzky_book_2001}. Mobile localizations, in other terminologies called
wave-fragments and gliders, are essential attributed of complexity. As you can see on Fig.~\ref{fig:sequence}, the rule is a distant anolog of 
replicator rules. 

\begin{figure}
\centering
\subfigure[T=1]{\includegraphics[width=0.15\textwidth]{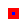}}
\subfigure[T=2]{\includegraphics[width=0.15\textwidth]{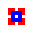}}
\subfigure[T=3]{\includegraphics[width=0.15\textwidth]{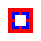}}
\subfigure[T=4]{\includegraphics[width=0.15\textwidth]{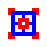}}
\subfigure[T=5]{\includegraphics[width=0.15\textwidth]{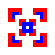}}
\subfigure[T=6]{\includegraphics[width=0.15\textwidth]{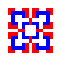}}
\subfigure[T=7]{\includegraphics[width=0.15\textwidth]{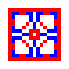}}
\subfigure[T=8]{\includegraphics[width=0.15\textwidth]{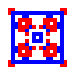}}
\subfigure[T=9]{\includegraphics[width=0.15\textwidth]{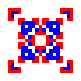}}
\subfigure[T=10]{\includegraphics[width=0.15\textwidth]{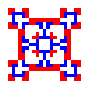}}
\subfigure[T=11]{\includegraphics[width=0.15\textwidth]{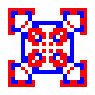}}
\subfigure[T=12]{\includegraphics[width=0.15\textwidth]{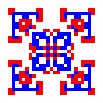}}
\subfigure[T=13]{\includegraphics[width=0.15\textwidth]{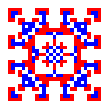}}
\subfigure[T=14]{\includegraphics[width=0.15\textwidth]{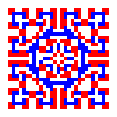}}
\subfigure[T=15]{\includegraphics[width=0.15\textwidth]{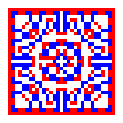}}
\caption{Configurations of defensive inhibition rule CA developed from 
an excited singleton. Resting cells are white, excited
cells are red and refractory cells are blue colored.} 
\label{fig:sequence}
\end{figure}

A singular initial excitation leads for formation of four mobile localizations (wave-fragments)
traveling South-West, South-West, North-West and North-East (Fig.~\ref{fig:sequence}de). These localizations
however split onto two localizations each (Fig.~\ref{fig:sequence}f) merge together (Fig.~\ref{fig:sequence}g) but 
then annihilate almost everywhere but in concavities of the growing pattern (Fig.~\ref{fig:sequence}h). The remained 
domains of excitation produce new mobile localization and the process of replication and annihilation continues.

\begin{figure}
\includegraphics[width=0.49\textwidth]{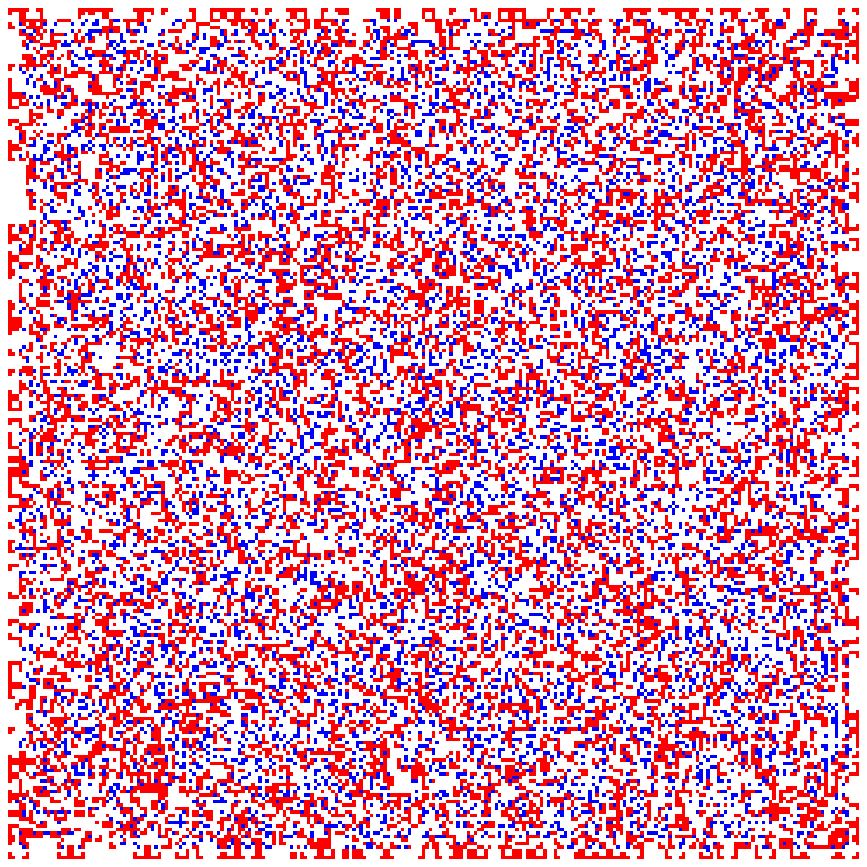}
\includegraphics[width=0.49\textwidth]{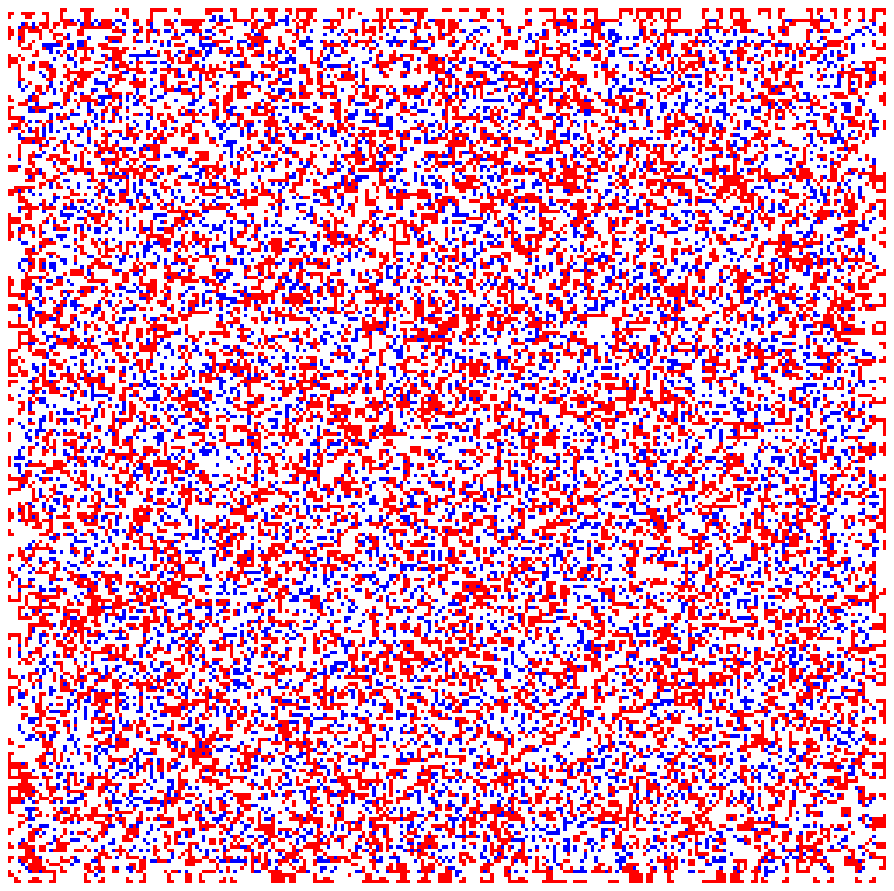}
\caption{Snapshots of $300 \times 300$-cell automaton governed by defensive inhibition rule.
(a)~development from excited singleton, configuration at step $T=119$, 
(b)~development from random configuration of uniformly distributed states, sites in the centered sub-lattice of 
$200 \times 200$ cells were assigned values at random, configurastion recorded at step $T=72$.} 
\label{fig:withtime}
\end{figure}

Eventually, spatio-temporal dynamics becomes so complicated that large-scale patterns emerged from excited singleton
look -- by naked eye -- as disordered as configuration developed from initial random configuration (Fig.~\ref{fig:withtime}). 
These patters are comprised of frequently born mobile localizations, gliders, which collide with each other and produce new 
localization. The frequency of their reproduction is however so high and the localization's life-time is so short that it 
is impossible to trace them visually.

\section{Cellular Automata with Memory}

Standard CA are ahistoric (memoryless): the transition function depends on the cells neighborhood configuration only at the preceding time step. Historic memory can be embedded in CA by presenting every cell by a mapping of its states in the previous time steps. 
Namely, we propose to maintain the transition rules $(\phi)$ unaltered, but to make them acting on the cells featured by a 
function of their previous states: $\sigma^{(T+1)}_{i} = \phi \Big(s^{(T)}_{j}\in \mathcal{N}_{j}\Big)$, where
$s^{(T)}_{i}$ is a state function of the series of states of the cell $i$ up to time-step $T$. 
Cellular automata implementing memory in cells will be termed \textit{historic}, and the standard ones \textit{ahistoric}.

Thus, cells can be featured by the most frequent of their three last states~\cite{RAS1}: 
$$s^{(T)}_{i}= mode\{\sigma^{(T-2)}_{i}, \sigma^{(T-1)}_{i}, \sigma^{(T)}_{i}\} .$$ 
In case of a tie, i.e. the three states 
present once, the last state will feature the cell. The \textit{historic} and
 \textit{ahistoric} patterns coincide up to $T=3$, with the initial assignations 
$s^{(1)}_{i}=\sigma^{(1)}_{i}$, $s^{(2)}_{i}=\sigma^{(2)}_{i}$. Memory becomes operative after $T=3$ .

Figure~\ref{fig:fig3} shows the effect of mode memory starting from a simple configuration. 
As a general rule, memory tends to restrain the evolution as shown in the case of mode memory in 
Fig.\ref{fig:fig3}. It is generally so from the beginning of the effective memory action, so at $T=3$ the 
outer excited cells in the actual pattern evolution are not featured as excited but as resting cells, as this 
is the their most frequent state up to this time step (twice resting versus one excited). Typically, the 
series of evolving patterns with memory diverges from the ahistoric evolution already at $T=4$. From this 
early time-step, the  patterns  with memory turn out less expanded, as shown in Fig.\ref{fig:fig3}. The restraing 
effect of memory may result fatal starting with simple configurations such as the singleton of  
Fig.\ref{fig:sequence}, which extinguishes at $T=4$ with mode memory \renewcommand{\baselinestretch}{0.0}  \footnote{
The only cells not featured as resting after $T=3$ are those refractory at $T=3$:\hspace{-0.1 cm} 
\begin{tabular}{c@{}c@{}c@{}c@{}c@{}c@{}}
\w \w \x \w \w\\
\w \x \x \x \w\\
\x \x \x \x \x\\
\w \x \x \x \w\\
\w \w \x \w \w\\
\end{tabular}~, which implyes immediate extinction.}  \renewcommand{\baselinestretch}{1.5}  
, a fact that led us to change the simple initital configuration
to demostration purposes to that of Fig.~\ref{fig:fig3}.

\begin{figure}
\hspace{0.7 cm}\includegraphics [width=0.88\textwidth]{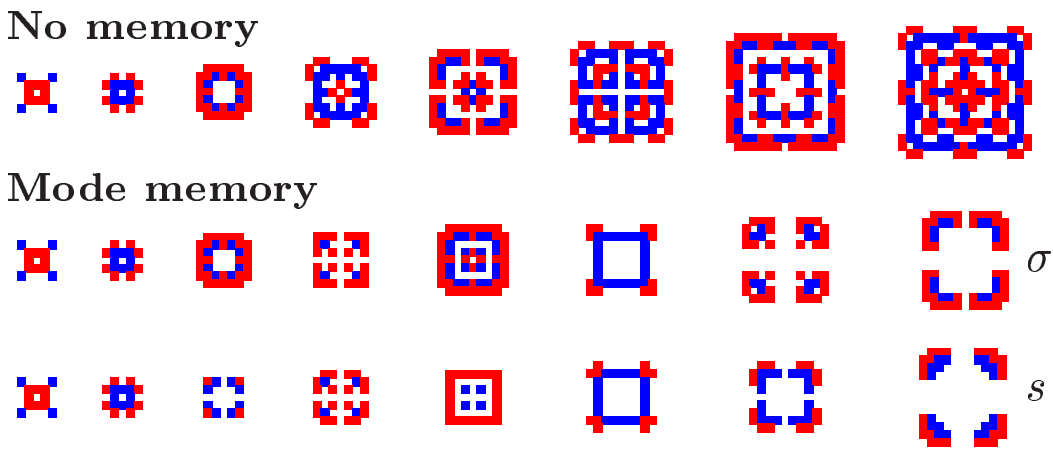}
\caption{Effect of mode memory starting from a simple configuration. The 
last series of evolving patterns shows the underlying patterns with cells featured by their most frequent 
state ($s$) along the last three time steps; these patterns generate the actual patterns ($\sigma$)
 with mode memory above-right them.}
\label{fig:fig3}
\end{figure}

Figure~\ref{fig:random} shows the effect of memory when initially every cell of the $100 \times 100$ cell sub-lattice of 
$300 \times 300$ cell lattice receive resting state with probability 0.8 and excited and refractory states with probability 0.1 each. 
The growing excitation pattern is `guided' by excited wave-fragments. There two immediately visible phenomenological 
differences between spatio-temporal dynamics of defensive inhibition CA with memory and without memory. First, excitation-inhibition pattern
in CA with memory spread two times slower compared to that of CA with memory. So, memory `delays' space-time developments. Second, 
configurations of CA without memory feature singletons of excited and refractory states, while in the configurations of CA with memory, 
more accentuated fiber-like clusters of excited and refractory states. There are also hints that small domains of labyrinthin-like patters of 
refractory states are surrounded by broken wave fronts of excitation (Fig.~\ref{fig:random}). 

\begin{figure}
\centering
\subfigure[$T=30$]{\includegraphics[width=0.31\textwidth]{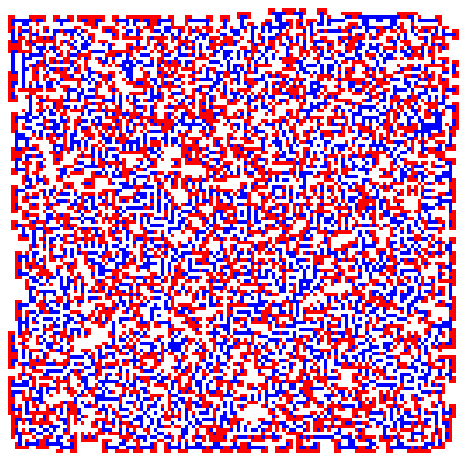}}
\subfigure[$T=60$]{\includegraphics[width=0.31\textwidth]{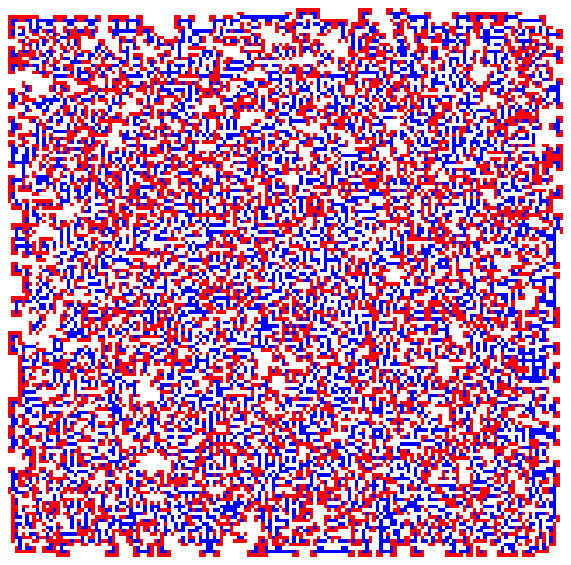}}
\subfigure[$T=90$]{\includegraphics[width=0.31\textwidth]{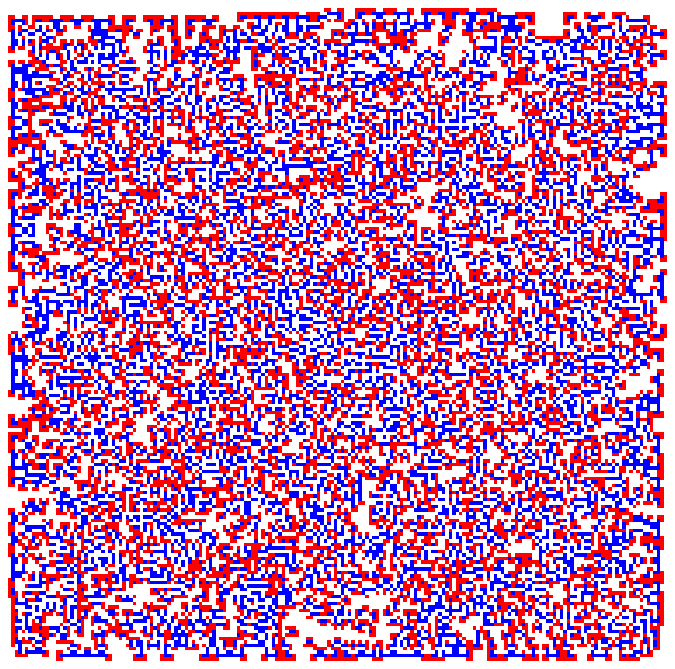}}
\subfigure[$T=120$]{\includegraphics[width=0.31\textwidth]{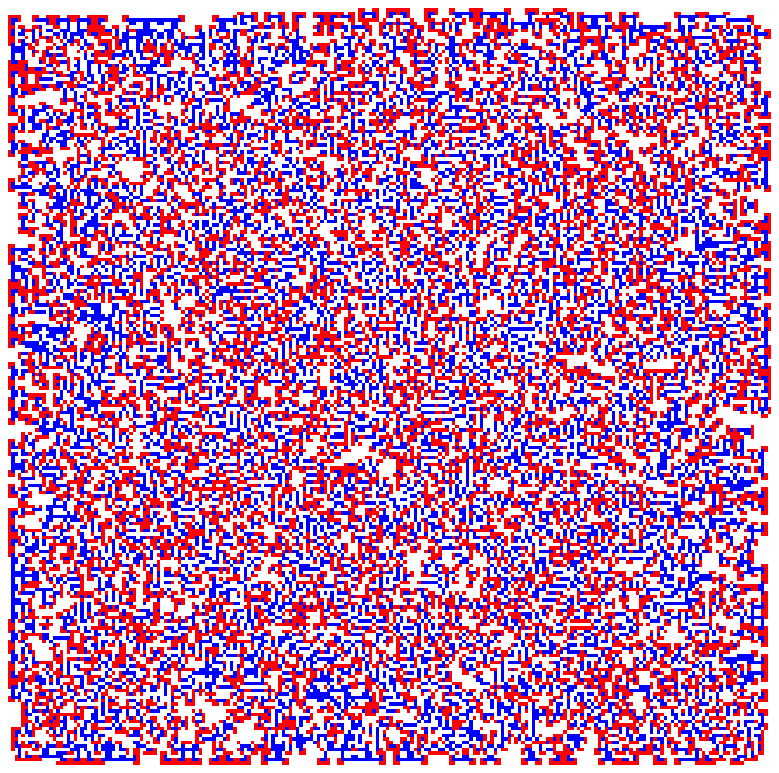}}
\subfigure[$T=150$]{\includegraphics[width=0.31\textwidth]{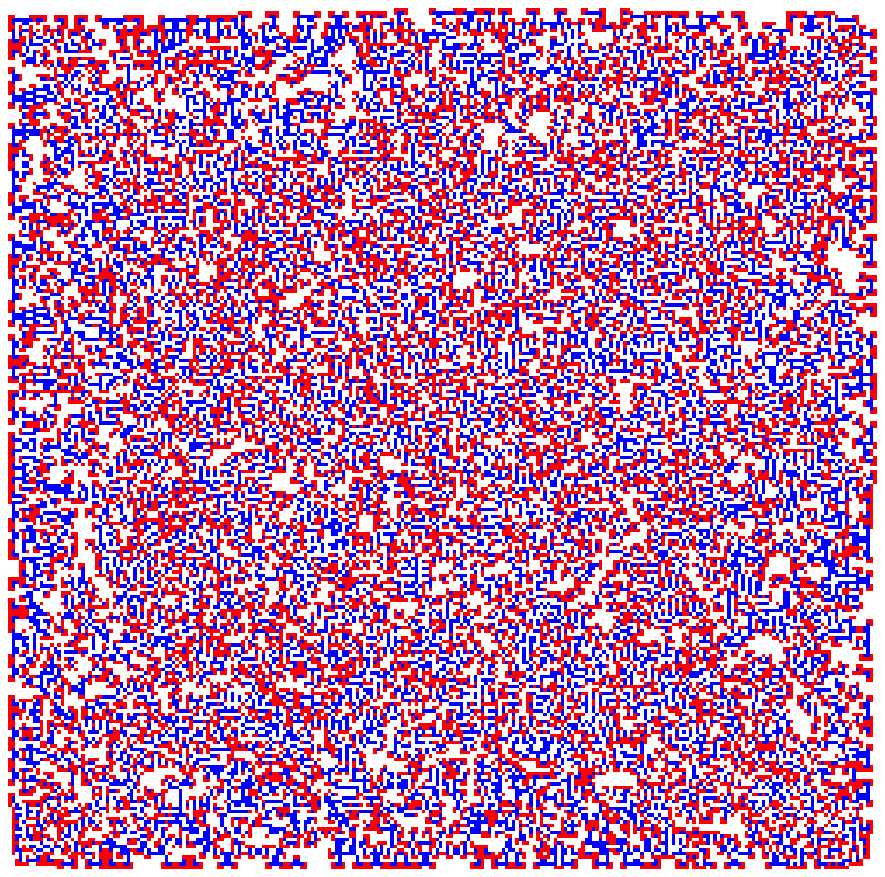}}
\subfigure[$T=180$]{\includegraphics[width=0.31\textwidth]{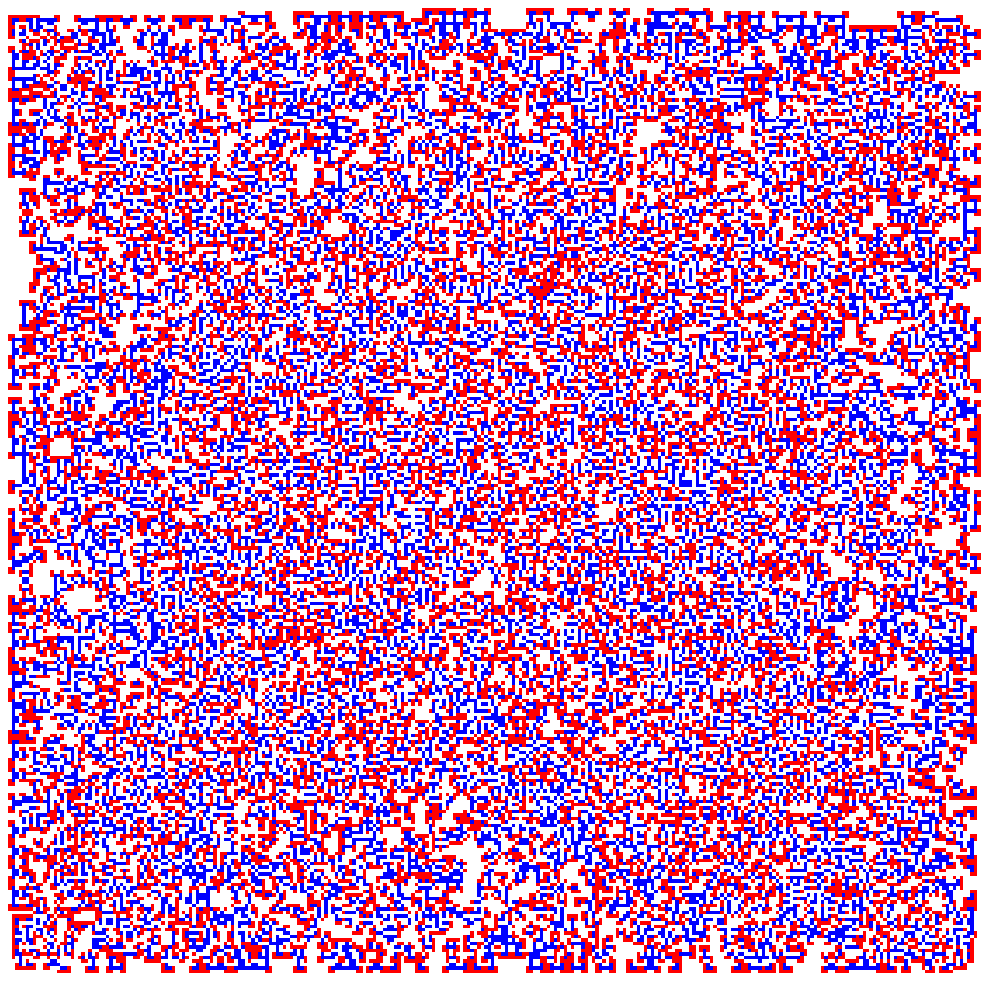}}
\caption{Development of CA with memory from random small configuration. }
\label{fig:random}
\end{figure}

It should be emphasized that the memory mechanism considered here is different from that of 
other CA with memory reported previously (see\cite{WOL2},p.118; \cite{ILA1},p.43; or 
class $\mathbb {CAM}$ \cite{ADA},p.7). Typically, higher-order-in-time rules incorporate memory into the
transition rule, determining the configuration at time $T+1$ in terms of the
configurations at previous time-steps. Thus, in second order in time (memory
of capacity two) rules, the transition rule operates as : $%
\sigma^{(T+1)}_{i}= \Phi \Big (\sigma^{(T)}_{j}\in \mathcal{N}_{i}, \sigma
^{(T-1)}_{j}\in \mathcal{N}_{i}\Big)$ . \textit{Double} memory (in transition rule and in cells) can be implemented as : $\sigma^{(T+1)}_{i}= \Phi\Big(s^{(T)}_{j}\in\mathcal{N}_{i},s^{(T-1)}_{j} \in\mathcal{N}_{i}\Big)$. Particularly interesting is the reversible formulation: $
\sigma^{(T+1)}_{i}= \phi\Big(\sigma^{(T)}_{j}\in \mathcal{N}_{i}\Big) \ominus\sigma^{(T-1)}_{i}$, reversed as $\sigma^{(T-1)}_{i}= \phi\Big(\sigma
^{(T)}_{j}\in \mathcal{N}_{i}\Big) \ominus \sigma^{(T+1)}_{i}$ . 
To preserve reversibility, the reversible formulation with memory must be : $\sigma^{(T+1)}_{i}= \phi \Big(s^{(T)}_{j} \in \mathcal{N}_{i}\Big)\ominus \sigma^{(T-1)}_{i}$ \cite{RAS5} .

Some authors, for example Wolf-Gladrow \cite{WOG}, define rules with \textit{%
memory} as those with dependence in $\phi $ on the state of the cell to be
updated. So elementary rules with no \textit{memory}, such as rule 90, take the
form : $\sigma^{(T+1)}_{i}=\phi\Big(\sigma^{(T)}_{i-1},\sigma ^{(T)}_{i+1}
\Big)$. Our use of the term memory is not this.

CA with memory in cells are cited by Wuensche and Lesser, \cite{WUE},p.15, who do not enter into  its study and state that ``CA with memory in cells would result in a qualitatively different behavior ".

	Let us point here that the mechanism of implementation of memory adopted in this work, keeping the transition rule unaltered but applying it to a function of previous  states, can be adopted in any dynamical system. In an earlier work
 (\cite{RAS2},\cite{RAS4},\cite{RAS1}) we explored the effect of embedding this kind of memory into \textit{discrete dynamical systems} : $x_{T+1}=f(x_{T})$ by means of $x_{T+1}=f(m_{T})$ with $m_{T}$ being a mean value of past states. We have studied this approach in, perhaps, the canonical example: the logistic map, which becomes with memory $x_{T+1}=m_{T}+\lambda m_{T}(1-m_{T})$. In \cite{RAS1}, we studied the effect of memory in a particular  Markovian stochastic processes (the random walk), 
$\textbf{p}^{\prime}_{T+1}=\textbf{p}^{\prime}_{T}\textbf{M}$ by means of 
$\textbf{p}^{\prime}_{T+1}={{\boldsymbol \pi}}^{\prime}_{T}\textbf{M}$ with 
${  {\boldsymbol \pi}_{T}  }$ being a weighted mean of the probability distributions up to $T$.

\section{\protect{Structurally Dynamic Cellular Automata}}

Structurally dynamic cellular automata (SDCA) were invented by Ilachinski and Halpern  \cite{ILA2}. 
They suggested the connections between the cells, are allowed to change according to rules similar 
in nature to the state transition rules associated with the conventional CA. This mean that given 
certain conditions, specified by the {\textit{link transition rules}}, links between rules may be 
created and destroyed. The  neighborhood  of  each  cell  is  now  becoming dynamic rather than fixed 
throughout the automaton.   Thus, state  and  link configurations of an SDCA are $both$ dynamic and  
are  continually altering each other.

In the Ilachinski and Halpern model, a SDCA consists of a  finite  set  of  binary-valued  $cells$ numbered  
1  to  $N$  whose connectivity is specified by a $N$$\times$$N$ connectivity matrix  in which $\lambda_{ij}$ 
equals 1 if cells $i$  and  $j$  are  connected;  0 otherwise. Thus, we have $\mathcal{N}^{(T)}_{i}=\{ j\big/ \lambda ^{(T)}_{ij}=1\}$  
and $\sigma^{(T+1)}_{i} = \phi\Big (\sigma^{(T)}_{j}\in \mathcal{N}^{(T)}_{i} \Big)$. The geodesic $distance$ 
between two cells $i$ and $j$, $\delta_{ij}$, is defined as the number of links in the shortest path between 
$i$ and $j$. We  say  that  $i$  and  $j$  are  \textit  {direct neighbors} if $\delta_{ij} = 1$, and that $i$ 
and $j$ are \textit {next-nearest neighbors} if $\delta_{ij} =2$, so $\mathcal{NN}^{(T)}_{i}=\{ j\big/ \delta^{(T)}_{ij}=2\}$. 
There are two types of link transition functions in SDCA: $couplers$ and $decouplers$, the former add new links, the later 
remove links. The set of coupler and decoupler determines the link transition rule:  
$\lambda^{(T+1)}_{ij}= \psi \Big(l^{(T)}_{ij}, \sigma^{(T)}_{i} , \sigma^{(T)}_{j} \Big)$ .

Instead of introducing the formalism of the SDCA, we deal here  with  just  
an example, in which  the  decoupler rule removes all links connected to cells in which both values are 
at refractory state ($\lambda ^{(T)}_{ij} = 1  \rightarrow \lambda^{(T+1)}_{ij}=0 $ ~$iff$~ $\sigma^{(T)}_{i}=2$ 
and $\sigma^{(T)}_{j}$=2) and the coupler  rule  adds links between all next-nearest neighbors sites in 
which both values are excited ($\lambda^{(T)}_{ij} = 0  \rightarrow \lambda^{(T+1)}_{ij} = 1$ ~$iff$~ $\sigma^{(T)}_{i}=1$ and $\sigma^{(T)}_{j}$ =1 and  $j\in \mathcal{NN}^{(T)}_{i}$). In the SDCA treated here, the transition rule for cell states 
is that of the generalized defensive inhibition rule, excitation interval [1,2]: resting cell is excited if
a ratio of  excited and connected to the cell neighbors to total number of connected neighbors lies in the 
the interval [1/8,2/8]. 

The initial scenario of Fig.\ref{fig:fig5} is that of Fig.\ref{fig:fig3} with the wiring network revealed, 
that of an Euclidean lattice with eight neighbors\renewcommand{\baselinestretch}{0.0}\footnote {
Moore neighborhood in CA terminology, in which, the generic cell
~\y~  has sixteen next-nearest neighbors : 
\begin{tabular}{c@{}c@{}c@{}c@{}c@{}c@{}c@{}c@{}c@{}c@{}c}
\n\n\n\n\n\\
\n\x\x\x\n\\
\n\x\y\x\n\\
\n\x\x\x\n\\
\n\n\n\n\n\\
\end{tabular}$\quad$.}\renewcommand{\baselinestretch}{1.5}. No decoupling is verified at the first iteration in 
Fig.\ref{fig:fig5}, but the excited cells generate new connections, most of them lost, together with some of the 
initial ones, at $T=3$. The excited cells at $T=3$ generate a $crown$ of new connections at $T=4$.

\begin{figure}[ht]
\begin{center}
\includegraphics [width=0.88\textwidth]{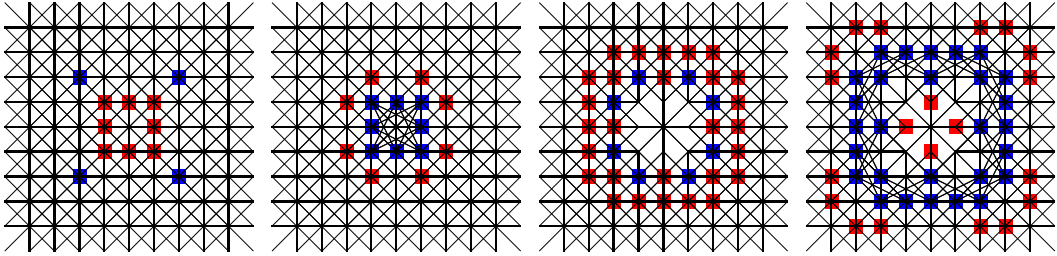}
\caption{The structurally dynamic cellular automaton described in Section 3, up to $T=4$. }
\label{fig:fig5}
\end{center}
\end{figure}

\section{\protect{A Structurally Dynamic Cellular Automaton with Memory}}

Memory can be embedded in links in a similar manner as in state values, so the link between any two cells is featured by a mapping of its previous link values: $ l_{ij}^{(T)}=l(\lambda _{ij}^{(1)},\ldots, \lambda _{ij}^{(T)})$. The $distance$ between two cells in a historic model ($d_{ij}$), is defined in terms of the $l$ instead of the $\lambda$ values,  so that $i$ and $j$ are  \textit  {direct neighbors} if $d_{ij} = 1$, and are \textit{next-nearest neighbors} if $d_{ij} =2$; $N_{i}^{(T)}=\{j/d_{ij}^{(T)}=1\}$,  and $NN^{(T)}_{i}=\{ j\big/ d^{(T)}_{ij}=2\}$ . In our approach here, the memory rule for links ($l$) is the same that of  state values ($s$). Generalizing the approach to embedded memory introduced in Sect.~2, the unchanged transition rules ($\phi$ and $\psi$) operate now on the featured link and cell state values:  
$\sigma ^{(T+1)}_{i} = \phi\Big(s^{(T)}_{j}\in N_{i} \Big)$ , $\lambda^{(T+1)}_{ij}= \psi \Big(l^{(T)}_{ij}, s^{(T)}_{i} , s^{(T)}_{j} \Big)$ \cite{RAS10},\cite{RAS11},\cite{RAS12}.
\par
Figure \ref{fig:fig6} shows the effect of mode memory in the scenario of Fig.\ref{fig:fig5}. Again,
memory has an $inertial$ effect. Also in what concerns to the link dynamics: the initial wiring network 
tends to be restored as made apparent in the underlying patterns of Fig.\ref{fig:fig6}. But inertial effect does not mean full restrain, as Fig.\ref{fig:fig7} shows. This figure, the ahistoric and mode
memory patterns at $T=20$, makes again apparent the preserving effect of memory .

\begin{figure}[ht]
\hspace{0.76 cm}\includegraphics[width=0.88\textwidth]{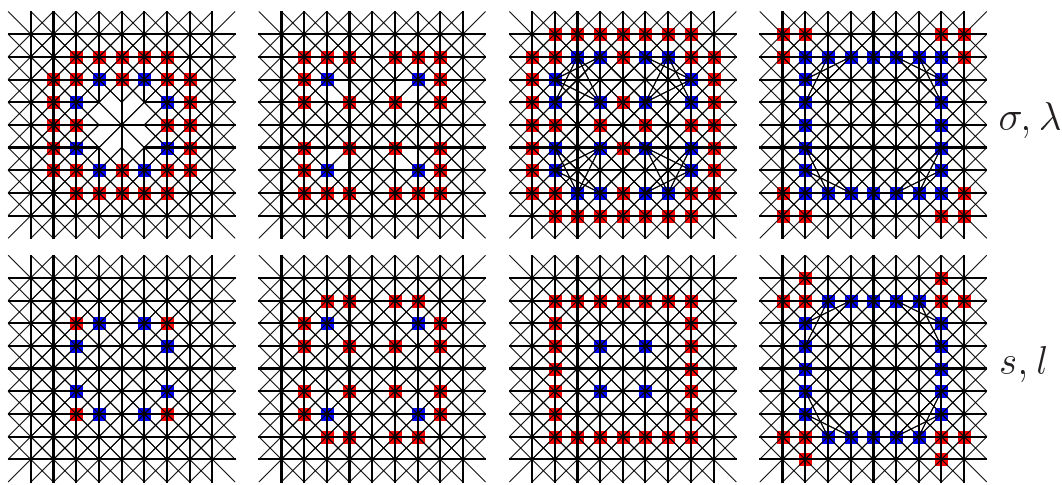}
\caption{The structurally dynamic cellular automaton with mode memory. Evolution from $T=3$ up to $T=6$ 
 starting as in Fig.\ref{fig:fig5}.  The first row of evolving patterns applies to the actual patterns, 
the second row shows the evolving patterns of the featuring states of cell and links.}\label{fig:fig6}
\end{figure}

\begin{figure}[ht]
\hspace{0.79 cm}\includegraphics[width=0.87\textwidth]{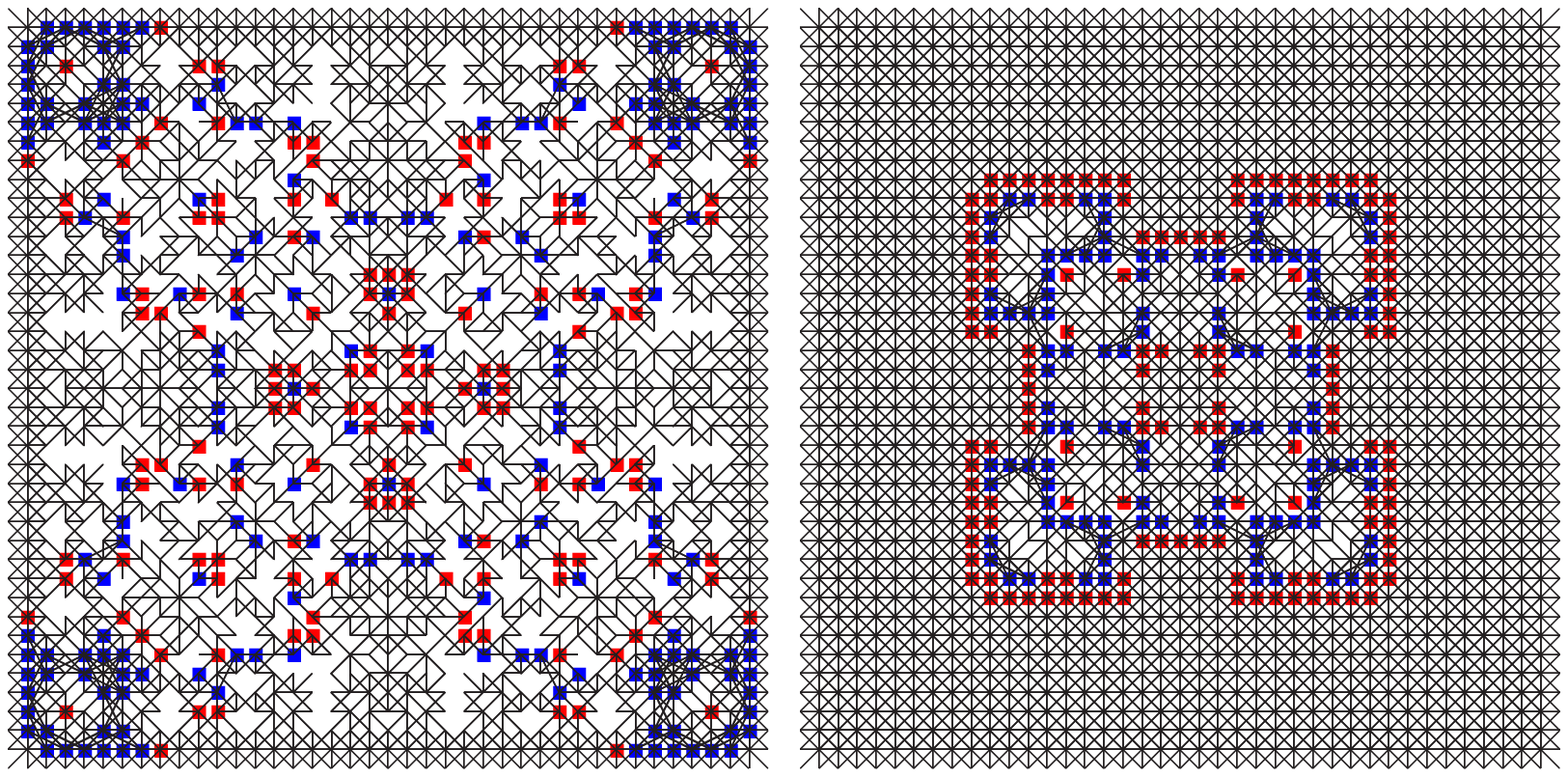}
\caption{The cellular automaton starting as in Fig.\ref{fig:fig5} at T=20, with no memory (left) 
and mode memory in both cell states and links.}\label{fig:fig7}
\end{figure}

\begin{figure}[ht]
\hspace{0.79 cm}\includegraphics[width=0.87\textwidth]{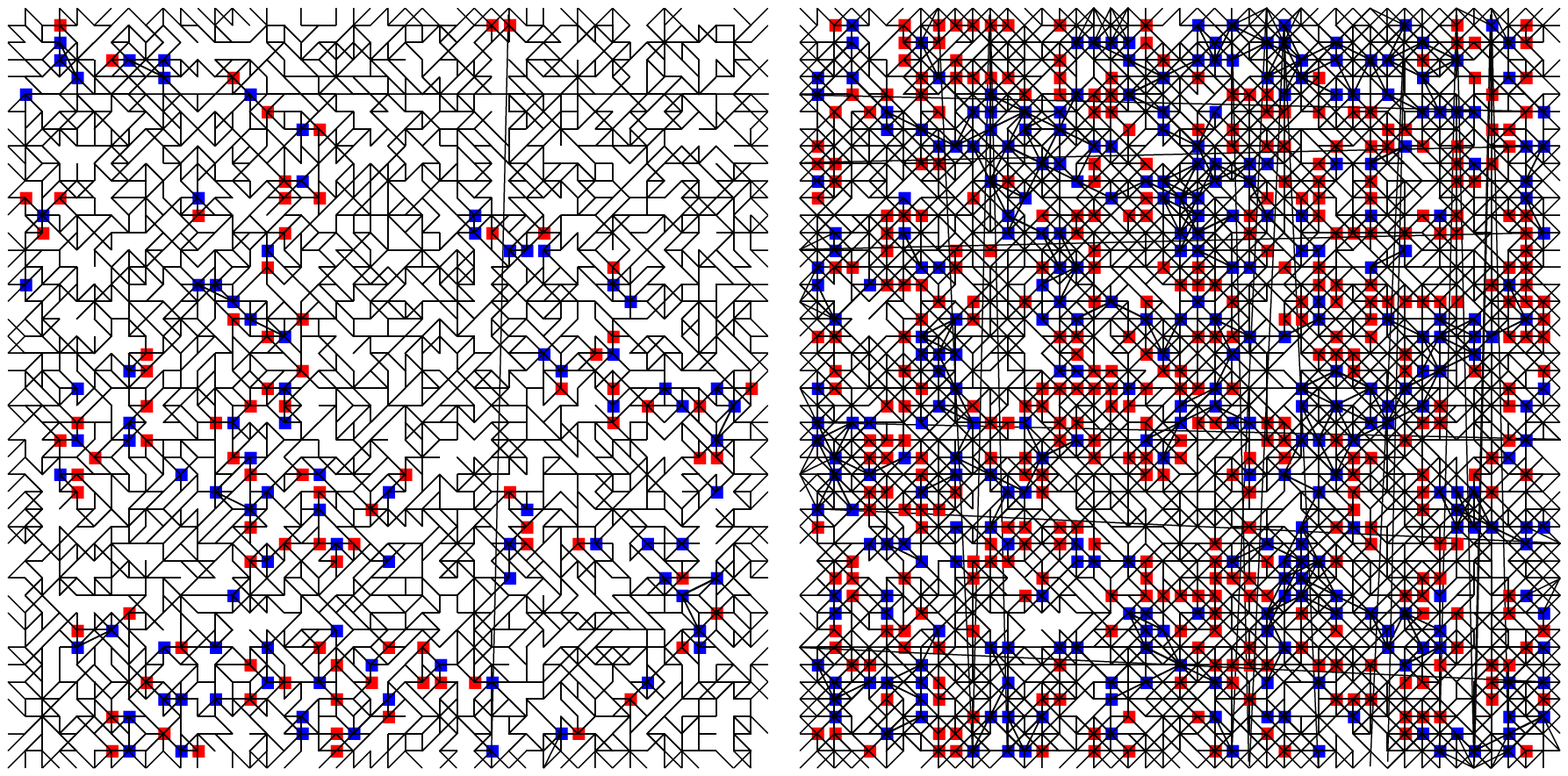}
\caption{The structurally dynamic cellular automaton starting at random in a $45\times45$ lattice. Patterns at
 T=50 with no memory (left), and mode memory.}
\label{fig:fig8}
\end{figure}


Figure~\ref{fig:fig8} shows the effect of mode memory at $T=20$ starting at random in what concerns to cell state levels
in an initially eight-neighbours lattice of size $45\times45$ with periodic boundary conditions. The wiring of the
border cells are not shown in Fig.\ref{fig:fig8} to avoid that the connections of the border cells (linked to their 
$opposite$ ones in the lattice) do mask the whole pattern. The ahistoric pattern shows a low density of non resting cells,
together with a irregular tessellation which clearly differs from the initial wiring network, but with a notable absence 
of links $traversing$ the pattern. Opposite, the historic pattern exhibits a higher presence of excited and refractory cells 
as well as higher density of links, somehow more reminiscent of the initial structure but much $traversed$ by connections
creeated during the evolution with memory.

Figure~\ref{fig:fig9} shows the evolution up to $T=50$ of the excited cell density, the average  number of
nearest neighbors  and next-nearest neighbors per site and the \textit{effective dimension}: the average 
ratio of  the number of next-nearest to nearest neighbors per site \footnote{Starting from an Euclidean lattice with eight neighbours, 
the effective dimension is 16/8=2.} (a discrete analogue to the continuous Hausdorff dimension), in the simulations of  Fig.~\ref{fig:fig8}. The evolution in the mode memory model shown in Fig.~\ref{fig:fig9} is that of the featured patterns
in all the parameters, and also that of the excited cell density and nearest neighbors of the actual patterns, corresponding
to the no dot marked curves.  Again the inertial effect of memory is shown in the evolution of 
excited cell density and  neighbors averages: the ahistoric curves (in red) stabilize in levels more distant (lower) to 
the initial ones than those of the ahistoric. The excited cell density evolution shows a fairly consistent parallelism 
between the actual (dotted) and featured patterns, whereas the actual and featured average number of nearest neighbors 
tend to approach, after an initial perturbation of the former. The general validity of the form of the curves  shown in
Fig.~\ref{fig:fig9} as well as the aspect of patterns at $T=20$ has been assessed by running ten more simulations with 
different random initial configurations.

\begin{figure}[ht]
\hspace{0.9 cm}\includegraphics[width=0.78\textwidth]{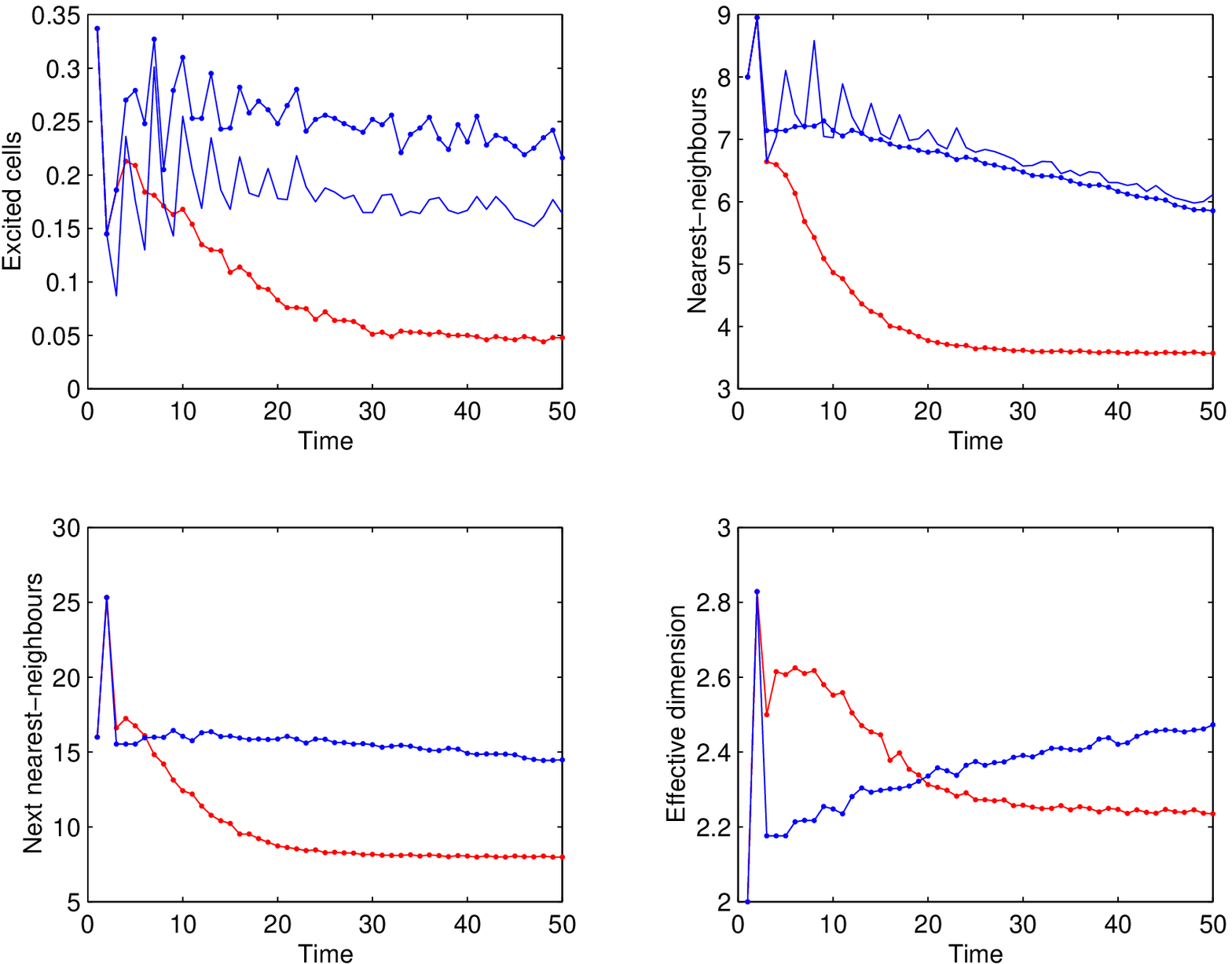}
\caption{Evolution of excited cell density, average number of nearest neighbours and next-nearest neighbours per site, and 
effective dimension in the simulation of Fig.\ref{fig:fig8}. Ahistoric simulation in red, mode memory one in blue.}
\label{fig:fig9}
\end{figure}

\section{\protect{Other memories}}

Memory can be implemented only in the cell state dynamics but not in the link dynamics, or viceversa.
Figure \ref{fig:fig10} shows the effect of such partial memory implementations starting as in Fig.{\ref{fig:fig5}}. 
Memory in cells appears much more determinant in preserving the initial features than memory in links.

\begin{figure}
\hspace{0.8 cm}\includegraphics[width=0.85\textwidth]{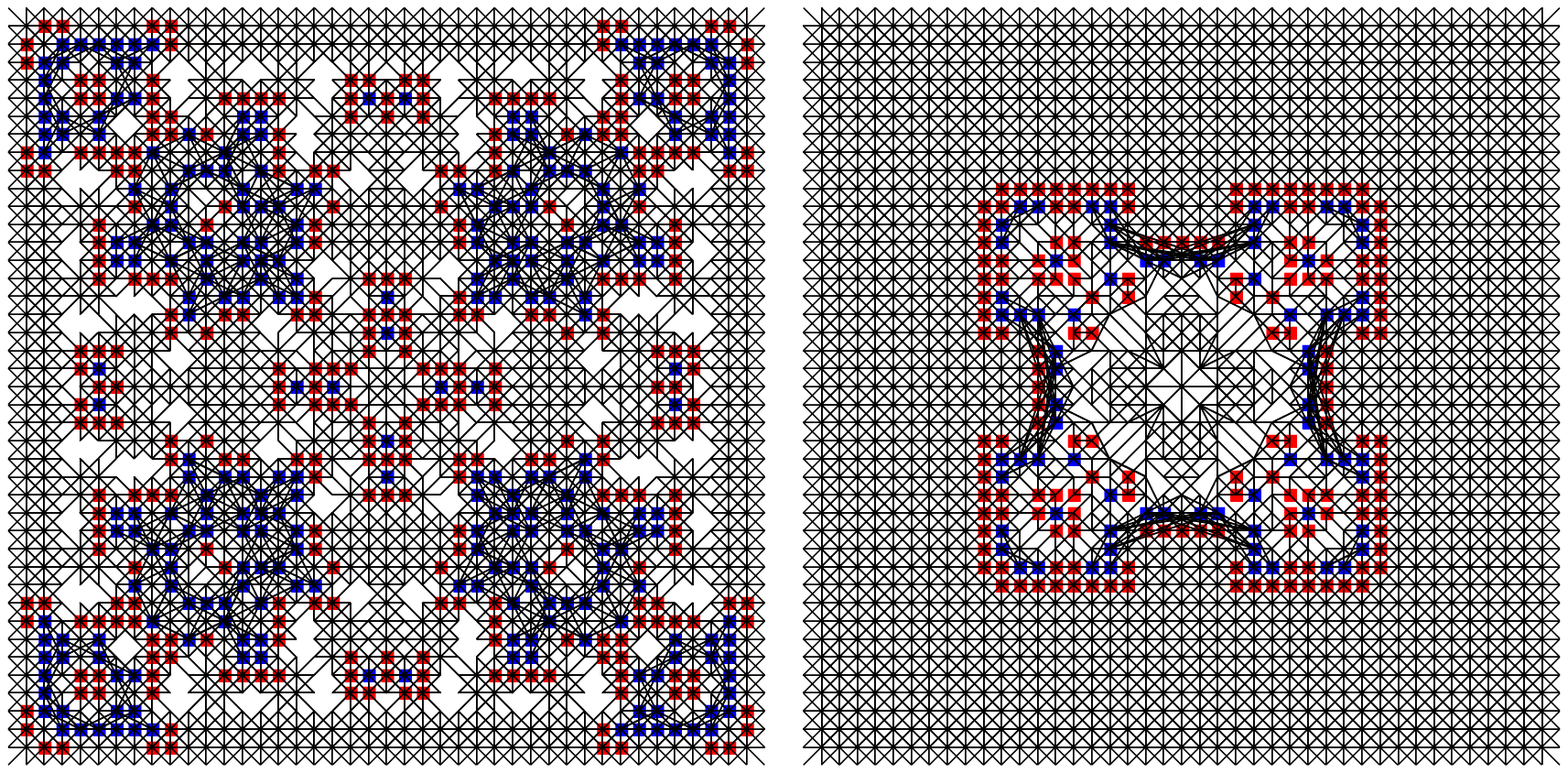}
\caption{The cellular automaton starting with memory only in links (left), and only in cell states.
Patterns at $T=20$ starting as in Fig.{\ref{fig:fig5}}. } 
\label{fig:fig10}
\end{figure}

An intermediate way of implementing updating will be that of adopting different 
time scales:  cell states are updated every time step, while cell links are 
updated every 10th or even every 100th time steps. This will be an analogue of 
fast and slow variables in non-linear equations. The idea may be translated to 
the way in which memory is implemented, which is planned for a future research work.

Memory of previous configurations can be embedded in the CA dynamics by means of 
a weighted average of previous states. Thus, a weighted average
memory  of $all$ previous states can be implemented as:  

$s^{(T)}_{i} =\left \{  
\begin{array}{lll}\vspace {0.1 cm}
         2                & \hbox{$if$} & m^{(T)}_{i} > 1.5\\ \vspace {0.1 cm}
         \sigma^{(T)}_{i} & \hbox{$if$} & m^{(T)}_{i} = 1.5\\ \vspace {0.1 cm}
         1                & \hbox{$if$} & 0.5 < m^{(T)}_{i} < 1.5 \\ \vspace {0.1 cm}
         \sigma^{(T)}_{i} & \hbox{$if$} & m^{(T)}_{i} = 0.5\\\vspace {0.1 cm}
         0                & \hbox{$if$} & m^{(T)}_{i} < 0.5 \\
\end{array} \right. $,
after $\quad$
$m^{(T)}_{i}(\sigma^{(1)}_{i},\ldots,\sigma^{(T)}_{i})=
\frac
{ \displaystyle \sigma^{(T)}_{i}+ \displaystyle \sum^{T-1}_{t=1}\alpha^{T-t}\sigma^{(t)}_{i}  } 
{1 + \displaystyle \sum^{T-1}_{t=1}\alpha^{T-t}   }
$, and
$l^{(T)}_{ij} =\left \{  
\begin{array}{lll}\vspace {0.1 cm}
         1           &  \hbox{$if$} & a^{(T)}_{ij} > 0.5\\ \vspace {0.1 cm}
         \lambda^{(T)}_{ij} & \hbox{$if$} & a^{(T)}_{i} = 0.5\\\vspace {0.1 cm}
         0           &  \hbox{$if$} & a^{(T)}_{ij} < 0.5 \\
\end{array} \right. $, after $\quad$
$m^{(T)}_{ij}=\displaystyle \frac
{\lambda^{(T)}_{ij}+\displaystyle \sum^{T-1}_{t=1}\alpha^{T-t}\lambda^{(t)}_{ij}} 
{1 + \displaystyle \sum^{T-1}_{t=1}\alpha^{T-t}}$   .

The choice of the memory factor $\alpha$ simulates the long-term or remnant memory effect: the limit case
 $\alpha =1$ corresponds memory with equally weighted records
($full$ memory model, $mode$ if $k=2$), whereas $\alpha << 1$ intensifies the contribution
 of the most recent states and diminishes the contribution of the past ones ($short$ type
 memory). The choice $\alpha = 0$ leads to the ahistoric model. Coding states as $0,1,\ldots,k-1$,
$\alpha$-memory is effective  if $\alpha > \displaystyle \frac{1}{2(k-1)}$  \cite{RAS4}.

\begin{figure}
\hspace{0.75 cm}
\includegraphics [width=0.89\textwidth]{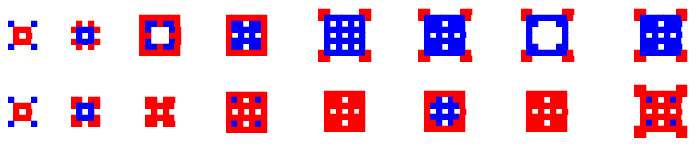}
\caption{The defensive inhibition rules with arithmetic mean memory of all previous states. Initial scenario of 
Fig.\ref{fig:fig3}. The first row of evolving patterns applies to the actual patterns, the second row shows the 
underlying patterns with cells featured by the rounded average state value of all its previous states.}\label{fig:fig11}
\end{figure}

As an example, Fig.\ref{fig:fig11} shows the effect of full memory
 $ m^{(T)}_{i}~=\displaystyle\frac{1}{T} \displaystyle \sum^{T}_{t=1}\sigma^{(t)}_{i} $ in the initial scenario of 
Fig.\ref{fig:fig3}.
Again, memory restrains the evolution of the defensive inhibition rule, but in a rather different way compared to the mode memory
in Fig.\ref{fig:fig3}. Embedding weighted memory typically has effect already at $T=2$, as  
in Fig.\ref{fig:fig11} by featuring the initially refractory cells (code 2) as excited (code 1) as being resting 
(code 0) at $T=2$ :   $1 = \displaystyle\frac{2+0}{2}$. This kind of, let us say, \emph{ad hoc} featuring when working
with more than three states is an inherent consequence of averaging, that tends to bias the result to the mean 
value. This led us to focus on a much more fair memory mechanism: the $mode$.

In Fig.\ref{fig:fig12} memory is implemented in an alternative way: cells are exciting if were excited at 
$T$ and/or at $T-1$. This implies a divergence in the evolving patterns from $T=3$ (as shown comparing 
Fig.\ref{fig:fig12} to Fig.~\ref{fig:fig3}), with a higher chance of resting cells to be in contact with $excited$ cells.

\begin{figure}\hspace{0.75 cm}
\includegraphics [width=0.89\textwidth] {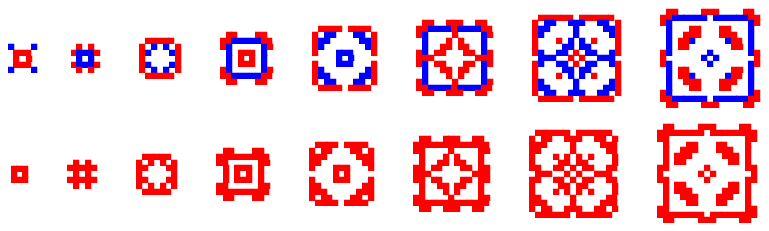}
\caption{The defensive inhibition rule featuring cells as exciting if excited at $T$ and/or at $T-1$. 
Initial scenario of Fig.~\ref{fig:fig3}. 
 The first row of evolving patterns applies to the actual patterns, the second row shows the 
cells as exciting if were excited at any of the two last time steps.}\label{fig:fig12}
\end{figure}

\section{\protect{Discussion}}

In present paper we undertook scoping studies of the effect of memory embedded in cells and links 
of specific structurally dynamic cellular automaton. We shown, that as a rule, memory notably alters the 
ahistoric dynamics. A complete analysis of the effect of memory on structurally dynamic cellular automata 
(SDCA) is left for future work which will develop a phenomenology of SDCA with memory, i.e. the full 
analysis of the rule space based on the morphological classification of patterns formed, the intrinsic 
parameters (e.g. Langton's lambda, Wuensche's Z parameter), the structure of global transition graphs 
(this would be feasible only in the 1D case), the entropy and other dynamics-related issues. 
Potential fractal features are also to come under scrutiny~\cite{SAN}.

Some critics may argue that memory is not in the realm of CA (or even of dynamic systems), but we believe 
that the subject is worth studying. At least CA with memory can be considered as a promising extension of 
the basic paradigm. A major impediment to modeling with CA stems from the difficulty of utilizing the CA 
complex behavior to exhibit a particular behavior or perform a particular function: embedding memory in 
states and links broadens the spectrum of CA as a tool for modeling. It is likely that in some contexts, a 
transition rule with memory could match the ``correct" behavior of the CA system of a given complex system 
(physical, biological, social and so on).

To substantiate this our initiative we tested how incorporating 
of the memory will influence dynamics of a system, more close to real life. We have to model
Belousov-Zhabotinsky (BZ) reaction, a classical example of excitable chemical system. Namely, we have chose to simulate
BZ reaction in sub-excitable mode: where local perturbations do not lead in general to target or spiral waves 
but to compact localized wave-fragments, or mobile localizations, 
propagating in the reaction space~\cite{sendina_2001}. 

We simulate the excitable chemical system using the two-variable Oregonator model~\cite{field_noyes_1974},\cite{tyson_fife}
modified to describe the light-sensitive analogue of the BZ reaction under applied illumination~\cite{beato_engel,krug}:
$$\frac{\partial u}{\partial t} = \frac{1}{\epsilon} (u - u^2 - (f v + \phi)\frac{u-q}{u+q}) + D_u \nabla^2 u$$
$$\frac{\partial v}{\partial t} = u - v$$
where variables $u$ and $v$ represent local concentrations of bromous acid HBrO$_2$ and the oxidized form of the
catalyst ruthenium Ru(III) respectively, $\epsilon$ sets up a ratio of time scale of variables $u$ and $v$, $q$ is a
scaling parameter depending on reaction rates, $f$ is a stoichiometric coefficient. Variable $\phi$ is a light-induced
bromide production rate proportional to the intensity of illumination and controls the excitability of the medium. 
The excitability is modified by light because the excited ruthenium catalyst reacts with bromo-malonic acid to produce 
bromide, which is an inhibitor of autocatalysis. The activator diffusion coefficient is $D_u$, whilst the diffusion term 
for $v$ is  absent as it is assumed that the catalyst is immobilized in a thin-layer of gel.
To integrate the system we used a Euler method with a five-node Laplacian operator, time step $\vartriangle t=10^{-3}$ and
grid point spacing $\vartriangle x = 0.25$, with the following parameters:
$\phi_0=0.076044$, $\epsilon=0.02272$, $f=1.4$, $q=0.002$. These parameters roughly
correspond to the region of higher excitability of the sub-excitable regime~\cite{sendina_2001}.

The waves were initiated by locally disturbing the initial concentration of species $u$, e.g. a few grid sites
in a chain are each given a value $u=1.0$; see configuration of initial perturbation in Fig.~\ref{fig:figbz}a. 
In memoryless model such a segment-perturbation generates  two localized wave fragments each, the wave-fragments 
travel a finite distance preserving their shape like quasi-particles or dissipative solitons (Fig.~\ref{fig:figbz}b). 

Analyzing various types of memory in the model we found that not only the memory function can boost or inhibit 
spreading wave-patterns but also distort topology of wave-activity. The following modes were considered:
\begin{itemize}
\item model $M_1$: $u^t=max(u^t,u^{t-1},u^{t-2})$ and $v^t=max(v^t,v^{t-1},v^{t-2})$,
\item model $M_2$: variable $v$ is not affected but  $u^t=max(u^t,u^{t-1},u^{t-2})$,
\item model $M_3$: variable $u$ is not affected but $v^t=max(v^t,v^{t-1},v^{t-2})$,
\item model $M_4$: $u^t=(u^t+u^{t-1}+u^{t-2})/3.$ and $v^t=(v^t+v^{t-1}+v^{t-2})/3.$,
\item model $M_5$: variable $u$ is not affected but $v^t=(v^t+v^{t-1}+v^{t-2})/3.$.
\end{itemize}

In the model $M_1$ wave-fragments, generated by initial perturbations start to expand. The same stimulus in the model 
$M_2$ lead to different outcomes, one local perturbation leads to formation of target wave, while another perturbation 
is transformed into  two expanding but still localized wave-fragments. All activity diminishes in model $M_3$. Both 
perturbations lead to target waves in models $M_4$ and $M_5$.

\begin{figure}
\centering
\subfigure[]{\includegraphics[width=0.2\textwidth]{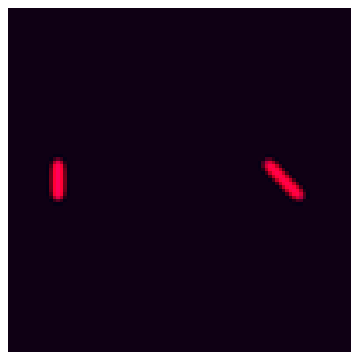}}
\subfigure[]{\includegraphics[width=0.2\textwidth]{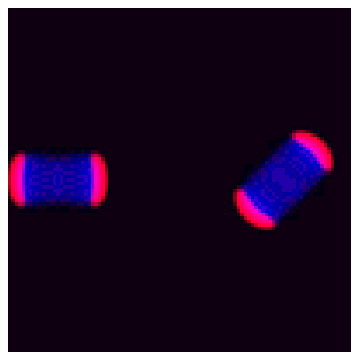}}
\subfigure[]{\includegraphics[width=0.2\textwidth]{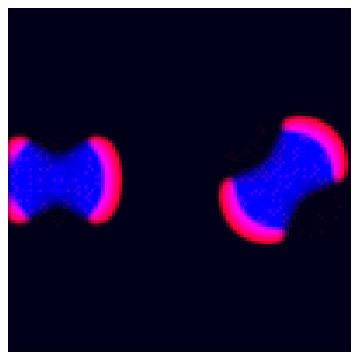}}
\subfigure[]{\includegraphics[width=0.2\textwidth]{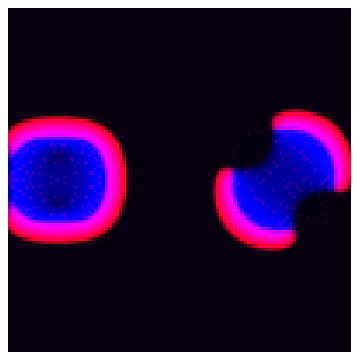}}
\subfigure[]{\includegraphics[width=0.2\textwidth]{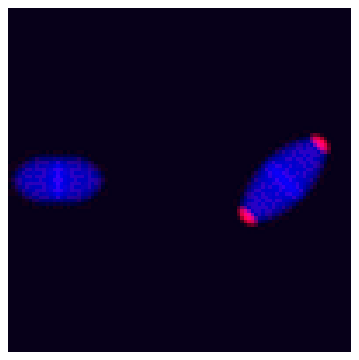}}
\subfigure[]{\includegraphics[width=0.2\textwidth]{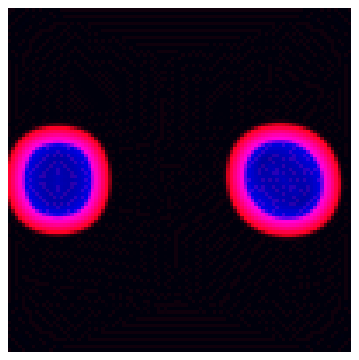}}
\subfigure[]{\includegraphics[width=0.2\textwidth]{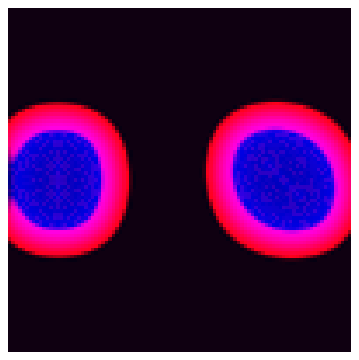}}
\caption{(a)~configuration of initial excitation-stimulation, 
(b)~localized wave-fragments in BZ model without memory, 
(c)~model $M_1$,
(d)~model $M_2$, 
(e)~model $M_3$,
(f)~model $M_4$, 
(g)~model $M_5$. }
\label{fig:figbz}
\end{figure}

The SDCA seems to be particularly appropriate for modeling the human brain function, -updating links between cells imitates variation of  synaptic connections between neurons represented by the cells -, in which the relevant role of memory is 
apparent. Models similar to SDCA have been adopted to build a dynamical network approach to quantum space-time physics~\cite{REQ}. 
Reversibility is an important issue at such a fundamental physics level; a generalization of the Fredkin's reversible construction 
is feasible in the SDCA scenario, which can be endowed with memory as:  
$\sigma^{(T+1)}_{i}= \phi \Big(s^{(T)}_{j} \in N^{(T)}_{i}\Big)\ominus \sigma^{(T-1)}_{i}$ , $\lambda^{(T+1)}_{ij}= \psi \Big(l^{(T)}_{ij}, s^{(T)}_{i} ,s^{(T)}_{j} \Big)\ominus \lambda^{(T-1)}_{ij}$  \cite{RAS26}. 
Technical applications of SDCA (and graph theory) may also be traced, so \cite{SASA} and \cite{SDCO}.

Anyway, besides their potential applications, SDCA with memory have an aesthetic and mathematical interest on their own. The study of the effect of memory on  CA has been rather neglected and there have been only limited investigations of SDCA  since its introduction in the late eighties \footnote {To the best of our knowledge, the relevant references on SDCA are \cite{HAL}--\cite{MAJ}, together with a review chapter in \cite{ILA1} and a section in \cite{ADA}.}. Nevertheless, it seems plausible that further study on SDCA (and Lattice Gas Automata with dynamical geometry \cite{LOV}) with memory \footnote{Not only in the basic paradigm scenario, but also in SDCA with random but value dependent rule transitions (which relates SDCA to Kauffman networks \cite{HAL}), and/or in SDCA with the extensions considered in \cite{MAJ}, such as unidirectional links.} should turn out to be profitable.

\section{Acknowledgement}
R. Alonso-Sanz work was supported by the Spanish Grant PR2006-0081.


\end{document}